\documentclass[aps,pre,preprint,groupedaddress,showpacs]{revtex4}
\usepackage{amsmath}
\usepackage{amssymb}
\usepackage{graphicx}

\begin{document}
\title{Non-Gaussian behaviour of a self-propelled particle on a substrate}%
\author{Borge ten Hagen}
 \affiliation{Institut f\"ur Theoretische Physik II, Weiche Materie,
Heinrich-Heine-Universit\"at D\"usseldorf, 
D-40225 D\"usseldorf, Germany}
\author{Sven van Teeffelen}
 \affiliation{Institut f\"ur Theoretische Physik II, Weiche Materie,
Heinrich-Heine-Universit\"at D\"usseldorf, 
D-40225 D\"usseldorf, Germany}
\author{Hartmut L\"owen}
\email[]{hlowen@thphy.uni-duesseldorf.de}
 \affiliation{Institut f\"ur Theoretische Physik II, Weiche Materie,
Heinrich-Heine-Universit\"at D\"usseldorf, 
D-40225 D\"usseldorf, Germany}
\date{\today}

\begin{abstract}
  The overdamped Brownian motion of a self-propelled particle which is
  driven by a projected internal force is studied by solving the
  Langevin equation analytically.  The ``active'' particle under study
  is restricted to move along a linear channel.  The direction of its
  internal force is orientationally diffusing on a unit circle in a
  plane perpendicular to the substrate. An additional time-dependent
  torque is acting on the internal force orientation. The model is
  relevant for active particles like catalytically driven Janus
  particles and bacteria moving on a substrate. Analytical results for
  the first four time-dependent displacement moments are presented and
  analysed for several special situations. For vanishing torque,
  there is a significant dynamical non-Gaussian behaviour at finite
  times $t$ as signalled by a non-vanishing normalized kurtosis in the
  particle displacement which approaches zero for long time with a
  $1/t$ long-time tail.
\end{abstract}

\keywords{Brownian dynamics, self-propelled particle, substrate, swimmer, active particles, diffusion} 
\pacs{82.70.Dd, 05.40.Jc}  

\maketitle

\section{Introduction}

The Brownian motion of self-propelled (``active'')
particles~\cite{Ramaswamy,review_RMP_Haenggi} bears much richer
physics than the traditional diffusive dynamics of passive particles.
Active particles can be modelled by moving under the action of an
internal force sometimes combined with an internal or external torque.
Realizations in nature are certain
bacteria~\cite{Berg:90,DiLuzio:05,Lauga:06,Hill:07,Shenoy:07} and
spermatozoa~\cite{Riedel:05,Woolley:03,Friedrich} which swim in
circles when confined to a surface~\cite{Ohta}.  In the colloidal
world, it is possible to prepare catalytically driven Janus
particles~\cite{Dreyfus:05,Dhar:06,Walther:08,Baraban:08} or biometric
particles~\cite{Fery:08} which perform self-propelled Brownian motion. For a recent investigation including confinement see~\cite{Popescu:09}.
On the macroscopic scale vibrated polar granular
rods~\cite{Kudrolli:07} on a planar substrate and even the
trajectories of completely blinded and ear-plugged
pedestrians~\cite{Obata:05} can be considered as rough realizations of
self-driven Brownian particles. If the particle is embedded in a
liquid (a ``swimmer''), as characteristic for colloids, the direction
of its driving force is fluctuating, in general, according to
orientational Brownian motion~\cite{Doi_Edwards_book,HL_Cyl,Klein}.
This gives rise to a non-ballistic translational motion of the
particles which is coupled to the fluctuating orientational degree of
freedom.

In most cases the direction of the self-propelling force is within the
plane of motion. For colloidal particles, however, it is possible to
confine the particle on a substrate by using, e.g., strong gravity
such that the particles are still freely
rotating~\cite{Baraban:08,Baraban2,Baraban_private} though they are
confined in a planar monolayer.  In this situation the component of
the self-propelling force which is normal to the surface is
compensated by the substrate, i.e., only the projection of the
self-propelling force onto the plane is driving the particle.
Therefore the translational motion is coupled to the (Brownian)
orientational motion~\cite{Teeffelen_PRE}.

In this paper we consider a one-dimensional model~\cite{Cates} for the
Brownian dynamics of a self-propelled particle on a substrate. The
particle is self-propelled along its orientational axis, which itself
is subjected to Brownian orientational diffusion. The particle is
confined to a channel, however, such that only the projected force in
channel direction is acting to drive the particle. The present study
is more general than earlier work in reference~\cite{Teeffelen_PRE}: first
of all, the present calculation resolves the Cartesian components of
the isotropic model on an unconfined plane.  Second, an arbitrary
time-dependence of the external torque is included here while this
torque was constant in~\cite{Teeffelen_PRE}. Finally, we calculate
time-dependent moments of the particle displacement up to fourth order
as compared to results up to second order in
reference~\cite{Teeffelen_PRE}.  The results are discussed for several
special cases.  In general, long-time self-diffusion is found.
Non-Gaussian behaviour is found for intermediate times as signalled in
the corresponding fourth cumulant.  The normalized kurtosis is
positive for small times, then changes sign and approaches zero from
below at long times with a $1/t$ long-time tail. This can be compared
to recent investigations for an undriven Brownian ellipsoid~\cite{Han:06}.
In the latter case, the kurtosis was found to be positive approaching
zero from above for long times with the same $1/t$ long-time tail.

This work represents a first step towards a many-body situation of
interacting self-propelled particles.  These are also realizable in
experiments (see, e.g.,~\cite{Dreyfus:05,Baraban:08,Kudrolli:07}).
The suitable theoretical framework is the many-body Smoluchowski
equation~\cite{Dhont_book}, from which one can derive a coupled
hierarchy of equations for the set of many-body distribution functions
similar in spirit to the traditional BBGKY
(Bogolyubov-Born-Green-Kirkwood-Yvon)
hierarchy~\cite{Bogolyubov1,Bogolyubov2,Uhlenbeck} for Liouville
dynamics, see also Felderhof~\cite{Felderhof} for a discussion in the
context of Brownian motion.  Therefore we think that this paper is
particularly appropriate for this issue dedicated to the 100th
anniversary of Prof.\ N.\ N.\ Bogolyubov.

This paper is organized as follows: In section \ref{Modell}, we propose and
motivate the model. The first four displacement moments 
are calculated analytically for the torque-free case in section \ref{-torque},
while section \ref{+torque} contains the results for a general time-dependent
torque.  Finally, in section \ref{conclusions}, we conclude and give an outlook on
possible future activities.
\section{The model}
\label{Modell}
\begin{figure}
\centerline{\includegraphics[width=0.65\textwidth]{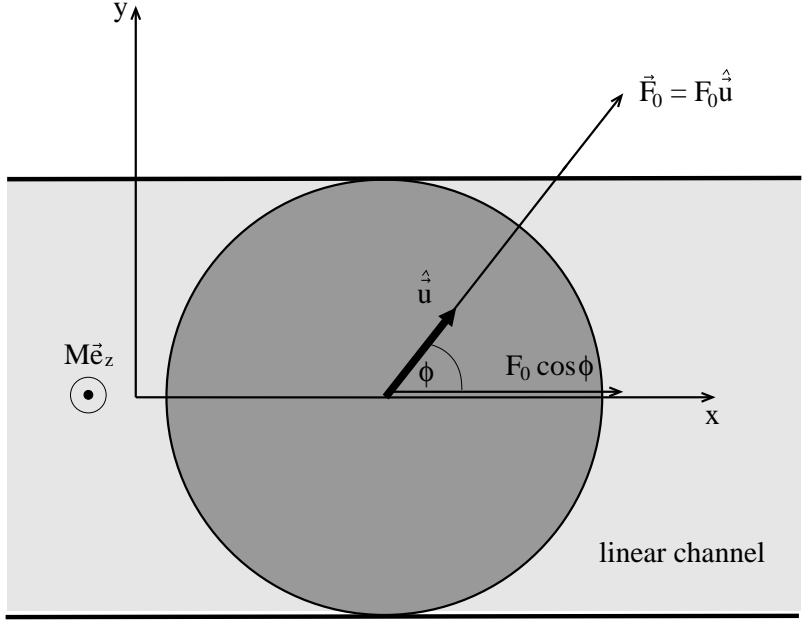}}
\caption{Sketch of the model system: A spherical colloidal particle
  (dark grey) is confined to a linear channel (light grey) along the
  $x$-direction. The self-propulsion is modelled by a constant
  effective force $\vec{F}_0$ along the particle orientation
  $\hat{\vec u}$. The latter is constrained to rotate in the
  $xy$-plane.  Only the projected force $F_0 \cos \phi$ drives the
  particle along the channel. A systematic, time-dependent torque
  $\vec M(t)=M(t)\hat{\vec e}_z$ is also indicated.}
\label{fig:modell}
\end{figure}
The model system under study consists of a self-propelled colloidal
sphere of radius $R$, which is confined to an infinite linear
channel in the $x$-direction, where it undergoes completely overdamped
Brownian motion (for a sketch see figure~\ref{fig:modell}). Whereas
the motion of the center-of-mass position $x$ is constrained to one
dimension, the orientation vector $\hat{\vec u}=(\cos\phi,\sin\phi,0)$
is constrained to rotate in the $xy$-plane. The self-propulsion of the
particle is modelled by a constant effective force along the particle
orientation $\vec F=F_0\hat{\vec u}$ and a generally time-dependent
effective torque in the $z$-direction $\vec{M}=M\hat{\vec e}_z$. Because
the particle is confined, only the projected force $\vec
F\cdot\hat{\vec e}_x=F_0\cos\phi\hat{\vec e}_x$ drives the particle
systematically along the channel.  Based on these considerations, the
translational and orientational motion is modelled by a Langevin
equation for the center-of-mass position $x$ and the orientation
vector $\hat{\vec u}$:
\begin{eqnarray}
\label{Langevinx1}
\frac{\mathrm{d}x}{\mathrm{d}t} &=& \beta D \left[F_0 \cos\phi + f(t)\right]\,,\\
\label{Langevinu1}
\frac{\mathrm{d}\hat{\vec u}}{\mathrm{d}t}&=&\beta D_\mathrm{r}\left[M(t)+g(t)\right]\hat{\vec e}_z\times \hat{\vec u}\,,
\end{eqnarray}
where $f(t)$ is a zero-mean, Gaussian white noise random force, which
is characterized by $\langle f(t)\rangle =0$ and $\langle f(t)
f(t')\rangle =2\delta(t-t')/(\beta^2 D)$, where angular brackets
denote a noise average. Correspondingly, $g(t)$ is a Gaussian white
noise random torque with $\langle g(t)\rangle =0$ and $\langle g(t)
g(t')\rangle =2\delta(t-t')/(\beta^2 D_\mathrm{r})$. Here, $\beta^{-1}=k_\mathrm{B}T$
denotes the thermal energy. $D$ and $D_\mathrm{r}$ are the translational and
rotational short-time diffusion constants, respectively. For a sphere
of radius $R$ in the three-dimensional bulk the two quantities fulfill
the relationship
\begin{equation}
\label{Konstanten}
\frac{D}{D_\mathrm{r}} = \frac{4 R^2}{3}\,. 
\end{equation}

Due to the constraint on the orientational motion, the vector
equation~(\ref{Langevinu1}) reduces to a Langevin equation for the
orientational angle $\phi$, which is given by
\begin{equation}
\label{Langevinphi1}
\frac{\mathrm{d}\phi}{\mathrm{d}t}=\beta D_\mathrm{r} \left[M(t)+g(t)\right]\,.
\end{equation}
If the initial time $t_0$ is set to be zero, the solutions of the Langevin equations (\ref{Langevinx1}) and
(\ref{Langevinphi1}) are given by
\begin{equation}
\label{phi1}
\phi(t)=\beta D_\mathrm{r}\int_{0}^t{\left[M(t')+g(t')\right]\mathrm{d}t'} + \phi_0
\end{equation}
and 
\begin{equation}
\label{x1}
x(t)=\beta D\left[F_0\int_{0}^t{\cos\phi(t')\mathrm{d}t'}+\int_{0}^t{f(t')\mathrm{d}t'}\right]+x_0
\end{equation}
with $\phi_0\equiv\phi(t_0)$ and $x_0\equiv x(t_0)$.  

The translation-rotation-coupling between these two equations, which
is due to the cosine in equations~(\ref{Langevinx1}) and (\ref{x1}), leads
to nontrivial results for the mean position $\langle x-x_0\rangle$ and
the mean square displacement $\langle (x-x_0)^2\rangle$ of the
particle position, as is shown in the following sections. Furthermore, the presence
of the coupling term leads to non-Gaussian behaviour, which is
reflected in a non-zero kurtosis. The latter is obtained by
calculating the fourth moment of the particle displacement
distribution further down.

We start our analysis in section~\ref{-torque} by studying the special
case of a vanishing systematic torque $M=0$. The more complex
situations of a constant torque $M(t)=M$ and a generally
time-dependent torque $M(t)$ are considered in section~\ref{+torque}.
\section{Results for a vanishing torque}
\label{-torque}
In this section, the simplest case with a vanishing torque is covered.
Solving equation~(\ref{Langevinphi1}) for $M(t)\equiv 0$ and averaging
gives
\begin{equation}
\label{Mittelwertphi1}
\left\langle \phi(t)\right\rangle =\phi_0
\end{equation}
and for the second moment 
\begin{equation}
\left\langle (\phi(t)-\phi_0)^2\right\rangle=2D_\mathrm{r}t.
\end{equation}
As $\phi(t)$ is a linear combination of Gaussian variables $g(t')$, according to Wick's theorem~\cite{Doi_Edwards_book}, $\phi(t)$ is Gaussian as well. Thus the probability distribution of $\phi$ proves to be 
\begin{equation}
\label{Verteilungphikonkret}
P(\phi ,t)=\frac{1}{\sqrt{4\pi D_\mathrm{r}t}}\exp\left({-\frac{(\phi-\phi_0)^2}{4D_\mathrm{r}t}}\right).
\end{equation}
Now the mean position of the particle can be calculated. From 
\begin{equation}
\label{Mittelwertcos}
\langle \cos\phi(t)\rangle  = \int_{-\infty}^\infty{\cos(\phi) P(\phi ,t) \mathrm{d}\phi}=e^{-D_\mathrm{r}t}\cos\phi_0 
\end{equation}
follows
\begin{equation}
\label{Moment11}
\langle x(t)-x_0 \rangle =\frac{4}{3}\beta F_0R^2 \cos(\phi_0)\left[1-e^{-D_\mathrm{r}t}\right],
\end{equation}
where we made use of equation (\ref{Konstanten}). Thus for short times one obtains 
\begin{equation}
\label{Mittelwerts}
\langle x(t)-x_0 \rangle = \frac{4}{3}\beta F_0R^2 \cos(\phi_0)D_\mathrm{r}t+\mathcal{O}\left(t^2\right)
\end{equation}
and for $t \gg D_\mathrm{r}^{-1}$ the $\phi_0$-dependent mean position converges towards 
\begin{equation}
\lim_{t \to \infty} \langle x(t)-x_0 \rangle =\frac{4}{3}\beta F_0R^2 \cos(\phi_0).
\end{equation}
\begin{figure}[tb]
\centerline{\includegraphics[width=0.65\textwidth]{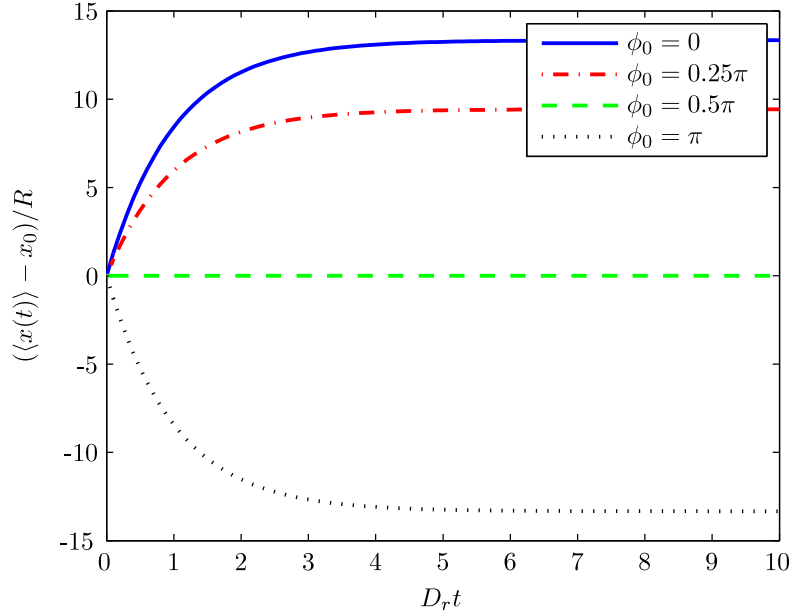}}
\caption{Mean position of a spherical particle without external torque for $\beta RF_0=10$ and different values of $\phi_0$.}
\label{fig:Moment11}
\end{figure}
The trajectory of the mean position $\langle x(t)\rangle$ is shown in figure~\ref{fig:Moment11} where
 the time $t$ is given in units of $D_\mathrm{r}^{-1}$, while the length $x$ is scaled by the particle radius $R$. 

To calculate the mean square displacement, the following integrals have to be solved: 
\begin{eqnarray}
\label{Moment2x1}
\left\langle (x(t)-x_0)^2\right\rangle & = & \beta^2 D^2\Bigl[F_0^2\int_0^t\mathrm{d}t_1\int_0^t\mathrm{d}t_2\langle \cos\phi(t_1)\cos\phi(t_2)\rangle  \nonumber \\ 
& & +2F_0\int_0^t\mathrm{d}t_1\int_0^t\mathrm{d}t_2\langle \cos\phi(t_1)f(t_2)\rangle  + \int_0^t\mathrm{d}t_1\int_0^t\mathrm{d}t_2\langle f(t_1)f(t_2)\rangle\Bigr].
\end{eqnarray}
The third summand can be calculated easily and equals $2t/(\beta^2D)$.
As $\langle \cos\phi(t)\rangle$ only depends on the random torque $g(t)$, $\langle \cos\phi(t)\rangle$ and $f(t)$ are statistically independent. Therefore the second summand vanishes. To calculate the first summand in equation~(\ref{Moment2x1}), the time correlation function is used. With $\phi_1\equiv \phi(t_1)$ and $\phi_2\equiv \phi(t_2)$ the required average can be written as
\begin{equation}
\label{Korrelation1}
\left\langle \cos\phi_1 \cos\phi_2\right\rangle_{t_1>t_2} = \int d\phi_1 \int d\phi_2 \cos\phi_1 \cos\phi_2 G(\phi_1,\phi_2, t_1-t_2)P(\phi_2, t_2)|_{t_1>t_2}.
\end{equation}
Here, $G(\phi_1,\phi_2, t_1-t_2)$ is the Green function, which is given by 
\begin{equation}
\label{Green1}
G(\phi_1,\phi_2 ,t_1-t_2)=\frac{1}{\sqrt{4\pi D_\mathrm{r}(t_1-t_2)}}\exp\left({-\frac{(\phi_1-\phi_2)^2}{4D_\mathrm{r}(t_1-t_2)}}\right).
\end{equation}
This yields 
\begin{equation}
\label{Korrelation2}
\left\langle \cos\phi_1 \cos\phi_2\right\rangle_{t_1>t_2} = \frac{1}{2}e^{-D_\mathrm{r}(t_1-t_2)}\left[1+\cos(2\phi_0)e^{-4D_\mathrm{r}t_2}\right].
\end{equation}
The expression for $\langle \cos\phi_1 \cos\phi_2\rangle_{t_2>t_1}$ is obtained in exactly the same way by replacing $t_1$ and $t_2$ with each other. Now, the first summand in formula (\ref{Moment2x1}) is calculated by simple integration and the mean square displacement can be written in the final form
\begin{equation}
\label{Verschiebungsquadrat1}
\left\langle (x(t)-x_0)^2\right\rangle = 2Dt + \left(\frac{4}{3} \beta F_0R^2\right)^2 \left[e^{-D_\mathrm{r}t}+D_\mathrm{r}t-1+\frac{1}{12} \cos(2\phi_0)\left(e^{-4D_\mathrm{r}t}-4e^{-D_\mathrm{r}t}+3\right)\right].
\end{equation}
\begin{figure}[tb]
\centerline{\includegraphics[width=0.65\textwidth]{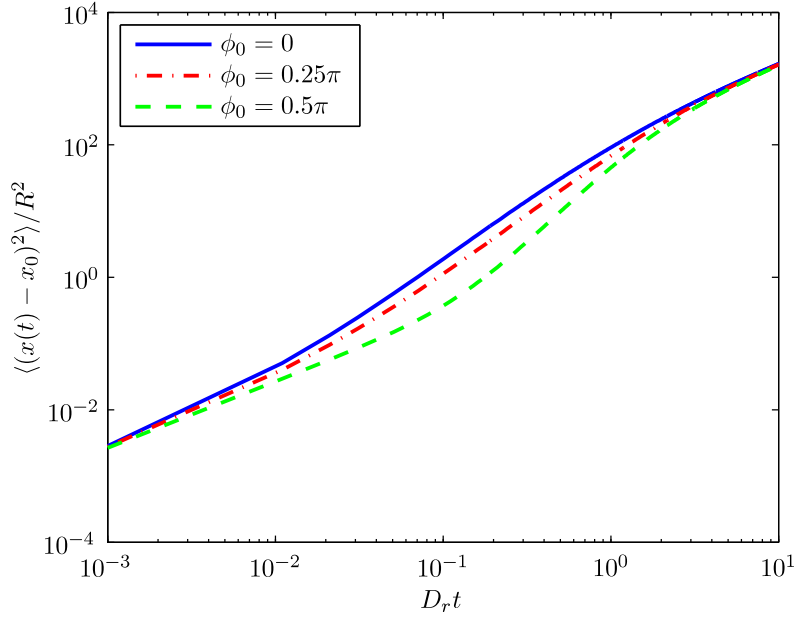}}
\caption{Mean square displacement of a spherical particle  for $\beta RF_0=10$ and different values of $\phi_0$.}
\label{fig:Moment21}
\end{figure}
The long-time diffusion coefficient $D_\mathrm{l}$ is given by
\begin{equation}
\label{Longtime1}
D_\mathrm{l} = \lim_{t \to \infty} \frac{1}{2t} \left\langle(x(t)-x_0)^2\right\rangle = D+\frac{8}{9}\left(\beta F_0R^2\right)^2 D_\mathrm{r}.
\end{equation}
Figure \ref{fig:Moment21} displays the results for the same cases that were examined in figure \ref{fig:Moment11}. The graph for $\phi_0=\pi$ coincides with the graph for $\phi_0=0$. As can be seen in the logarithmic plots and from the expression (\ref{Longtime1}), the initial angle $\phi_0$ is not relevant for times much longer than $D_\mathrm{r}^{-1}$. 

In the following the non-Gaussian behaviour  of the particle is investigated. For this purpose  skewness $S$ and kurtosis $\gamma$ are calculated. The non-Gaussian behaviour   is clearly signalled in the nonzero value of these quantities. In general, the skewness is given by
\begin{equation}
\label{Schiefe1}
S=\frac{\left\langle(x-\langle x\rangle)^3\right\rangle}{\left\langle(x-\langle x\rangle)^2\right\rangle^{3/2}}=\frac{\langle x^3\rangle -3\langle x\rangle \langle x^2\rangle +2\langle x\rangle ^3}{(\langle x^2 \rangle -\langle x\rangle^2 )^{3/2}},
\end{equation}
and the kurtosis is calculated as 
\begin{equation}
\gamma=\frac{\left\langle(x-\langle x\rangle)^4\right\rangle}{\left\langle(x-\langle x\rangle)^2\right\rangle^2}-3=\frac{\langle x^4\rangle -4\langle x\rangle \langle x^3\rangle +6\langle x\rangle^2 \langle x^2\rangle -3\langle x\rangle ^4}{(\langle x^2 \rangle -\langle x\rangle^2 )^2}-3. 
\end{equation}
For the third and fourth moments of $x$ -- in analogy to equation (\ref{Moment2x1}) -- one has to solve the integrals 
\begin{eqnarray}
\label{Moment3x1}
\left\langle (x(t)-x_0)^3\right\rangle & = & \beta^3 D^3\int_0^t\mathrm{d}t_1\int_0^t\mathrm{d}t_2\int_0^t\mathrm{d}t_3\Bigl[F_0^3\langle \cos\phi(t_1)\cos\phi(t_2)\cos\phi(t_3)\rangle  \nonumber \\ 
& & +3F_0\langle \cos\phi(t_1)\rangle \langle f(t_2)f(t_3) \rangle\Bigr]
\end{eqnarray}
and
\begin{eqnarray}
\label{Moment4x1}
\left\langle (x(t)-x_0)^4\right\rangle & = & \beta^4 D^4\int_0^t\mathrm{d}t_1\int_0^t\mathrm{d}t_2\int_0^t\mathrm{d}t_3\int_0^t\mathrm{d}t_4\Bigl[F_0^4\langle \cos\phi(t_1)\cos\phi(t_2)\cos\phi(t_3)\cos\phi(t_4)\rangle  \nonumber \\ 
& & +6F_0^2\langle \cos\phi(t_1)\cos\phi(t_2)\rangle \langle f(t_3)f(t_4) \rangle  + \langle f(t_1)f(t_2)f(t_3)f(t_4)\rangle\Bigr],
\end{eqnarray}
respectively. Before solving the time-integrals over the first summands, the time correlation functions 
\begin{eqnarray}
\label{Korrelation3}
\left\langle \cos\phi_1 \cos\phi_2\cos\phi_3\right\rangle_{t_1>t_2>t_3} &\!\!=\!\!& \frac{1}{2}\cos(\phi_0)e^{-D_\mathrm{r}(t_1-t_2+t_3)} \nonumber \\ &&+\frac{1}{4}e^{-D_\mathrm{r}(t_1+3t_2-4t_3)}\left[\cos(\phi_0)e^{-D_\mathrm{r}t_3}+\cos(3\phi_0)e^{-9D_\mathrm{r}t_3}\right]\qquad
\end{eqnarray}
and
\begin{eqnarray}
\label{Korrelation4}
&&\left\langle \cos\phi_1 \cos\phi_2\cos\phi_3\cos\phi_4 \right \rangle_{t_1>t_2>t_3>t_4} \nonumber \\
&= &\frac{1}{4}e^{-D_\mathrm{r}(t_1-t_2+t_3-t_4)}\left[1+\cos(2\phi_0)e^{-4D_\mathrm{r}t_4}\right] \nonumber \\ &&+\frac{1}{4}e^{-D_\mathrm{r}(t_1+3t_2-4t_3)}\Bigl\{\frac{1}{2}e^{-D_\mathrm{r}(t_3-t_4)}\left[1+\cos(2\phi_0)e^{-4D_\mathrm{r}t_4}\right] \nonumber \\
&&+\frac{1}{2}e^{-9D_\mathrm{r}(t_3-t_4)}\left[\cos(4\phi_0)e^{-16D_\mathrm{r}t_4}+\cos(2\phi_0)e^{-4D_\mathrm{r}t_4}\right]\Bigr\}\qquad
\end{eqnarray}
have to be evaluated. Here the notation $\phi_i \equiv \phi(t_i)$ with $i\in\{1,2,3,4\}$ is used again. 
Both in equation (\ref{Moment3x1}) and in equation (\ref{Moment4x1}), the remaining  terms can be calculated easily using the expressions already obtained for the first and second moments. The complete analytical results for the third and fourth moments
(and for the skewness and kurtosis) are presented in the appendix. 

\begin{figure}[htb]
\centerline{\includegraphics[width=0.65\textwidth]{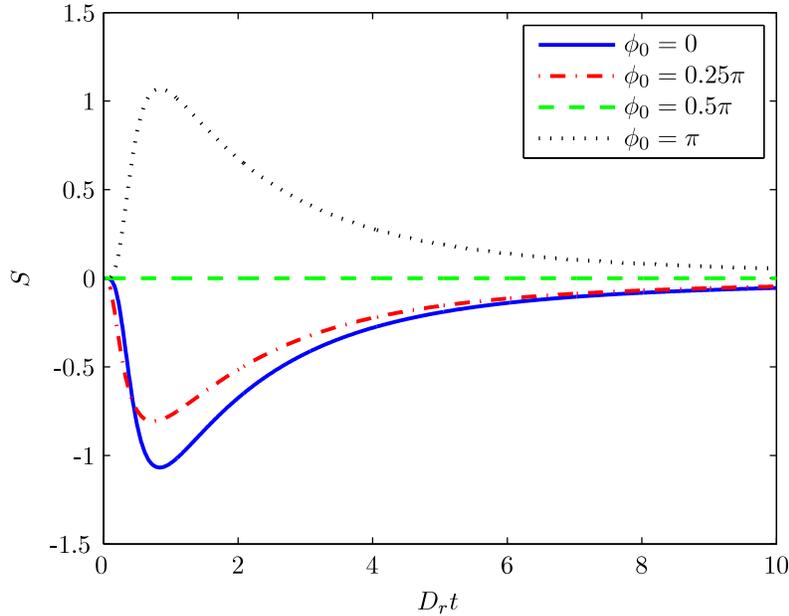}}
\caption{Skewness $S(t)$  for $\beta RF_0=10$ and different values of $\phi_0$.}
\label{fig:Schiefe}
\end{figure}
\begin{figure}[htb]
\centerline{\includegraphics[width=0.65\textwidth]{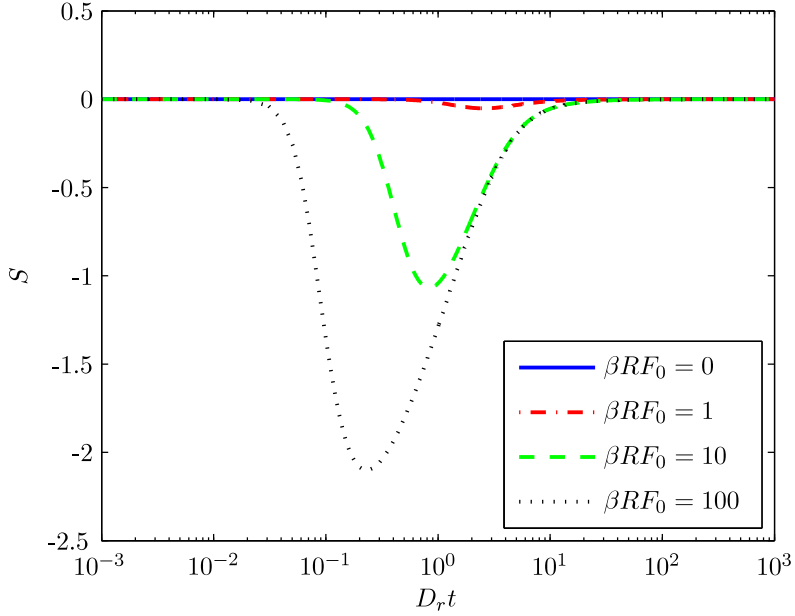}}
\caption{Skewness $S(t)$ for $\phi_0=0$ and different values of $\beta RF_0$.}
\label{fig:Schiefea}
\end{figure}
Figures \ref{fig:Schiefe} and \ref{fig:Schiefea} display the skewness $S$ of the probability distribution of the particle position for different values of the initial angle $\phi_0$ and the dimensionless quantity $\beta RF_0$ which determines whether the self-propulsion or the  motion due to the interaction with the solvent molecules is dominant. Figure~\ref{fig:Schiefe} shows that the sign of the skewness depends on $\phi_0$. If the $x$-component of the initial orientation is positive ($-0.5\pi<\phi_0<0.5\pi$), the skewness is negative, while initial angles between $0.5\pi$ and $1.5\pi$ lead to positive $S$. For symmetry reasons the skewness is zero for $\phi_0=0.5\pi$. Further analysis of formula (\ref{Schiefe}) (see the appendix) gives the leading long-time behaviour
of the skewness $S(t)$ as 
\begin{equation}
\label{Sl}
S(t) = \frac{8}{3}\,{\frac {{a}^{3}\cos \left( \phi_0 \right)  \left(  \left( \cos \left( 
\phi_0 \right)  \right) ^{2}-3 \right) }{ \left( 3+2\,{a}^{2} \right) 
\sqrt {4\,{a}^{2}+6}}}\,{(D_\mathrm{r}t)}^{-3/2} + \mathrm{o}\left(\frac{1}{t^{3/2}}\right), 
\end{equation}
where the abbreviation $a\equiv \beta RF_0$ is used, i.e., the skewness decreases proportionally to $t^{-3/2}$. 
\begin{figure}[htb]
\centerline{\includegraphics[width=0.65\textwidth]{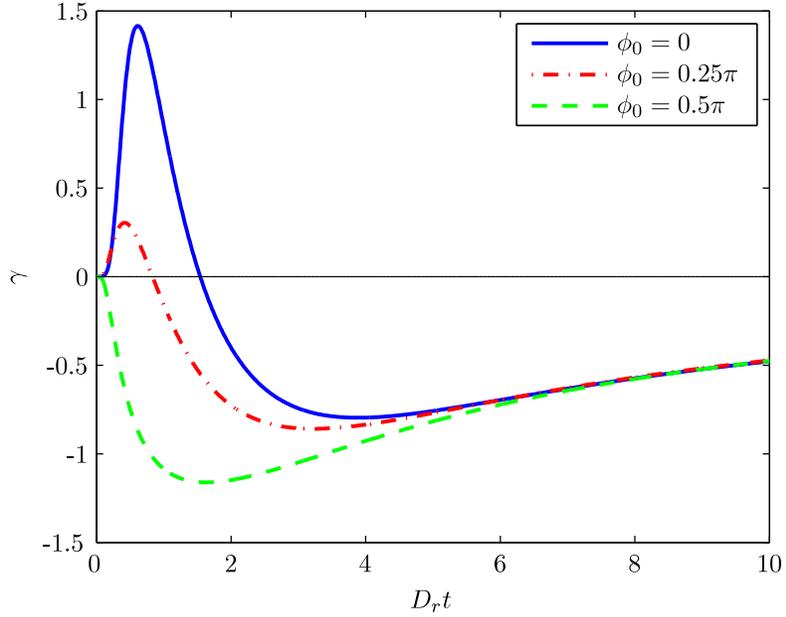}}
\caption{Kurtosis $\gamma(t)$  for $\beta RF_0=10$ and different values of $\phi_0$.}
\label{fig:Woelbung}
\end{figure}
\begin{figure}[htb]
\centerline{\includegraphics[width=0.65\textwidth]{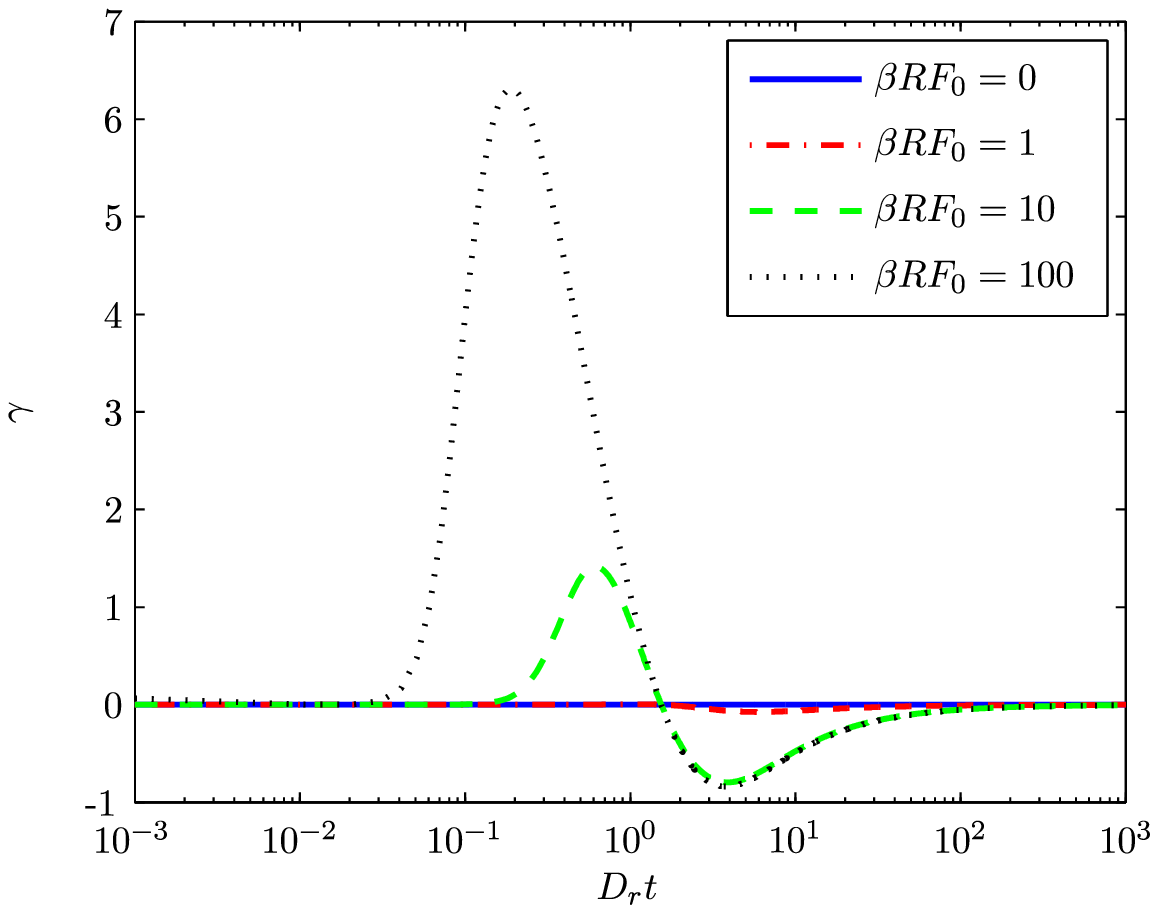}}
\caption{Kurtosis $\gamma(t)$  for $\phi_0=0$ and different values of $\beta RF_0$.}
\label{fig:Woelbunga}
\end{figure}
\begin{figure}[htb]
\centerline{\includegraphics[width=0.65\textwidth]{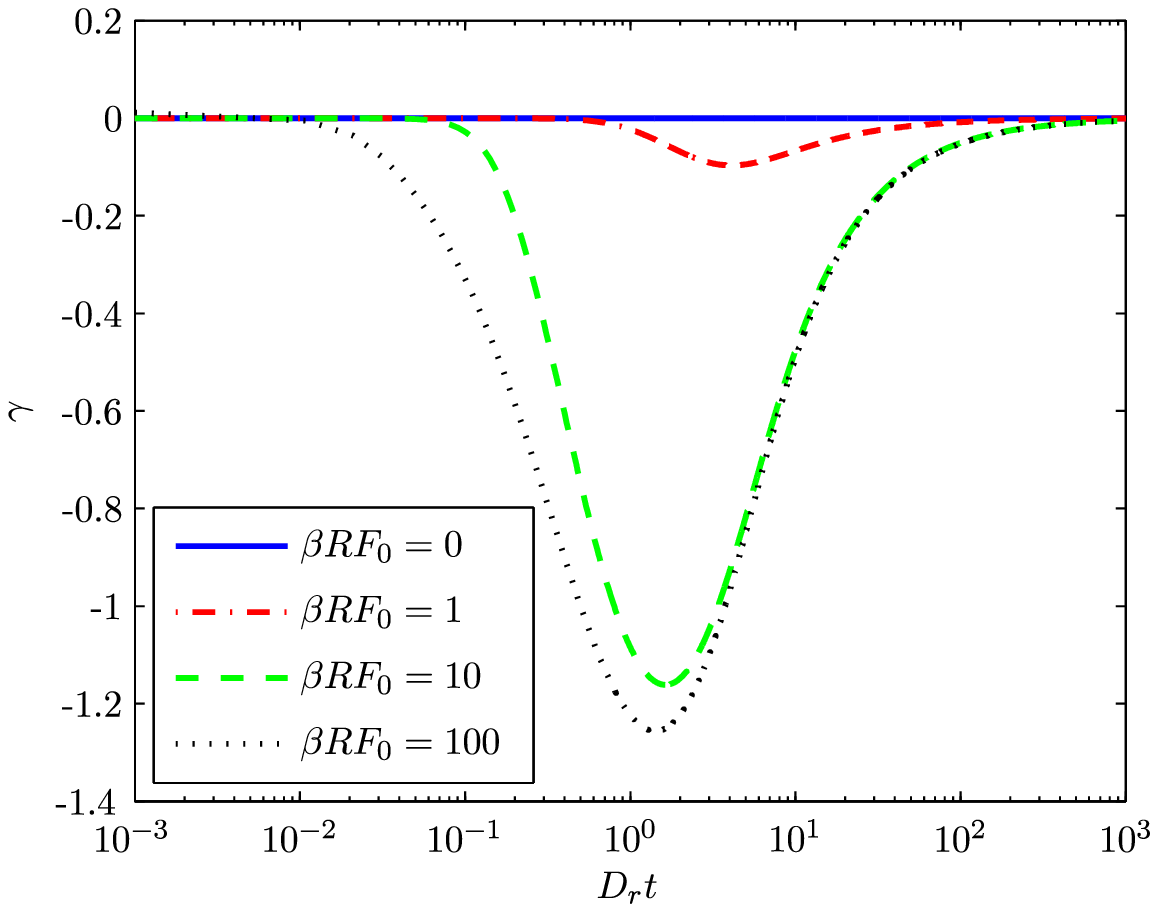}}
\caption{Kurtosis $\gamma(t)$  for $\phi_0=0.5\pi$ and different values of $\beta RF_0$.}
\label{fig:Woelbunga2}
\end{figure}
Similar analysis of formula (\ref{kurtosis}) for the kurtosis $\gamma (t)$ reveals a long time behaviour as
\begin{equation}
\gamma(t)={\frac{-21{a}^{4}}{9+12\,{a}^{2}+4\,{a}^{4}}}\,(D_\mathrm{r}t)^{-1}+\mathrm{o}\left(\frac{1}{t}\right).
\end{equation}
First of all, as can be seen from this formula and in figures \ref{fig:Woelbung}--\ref{fig:Woelbunga2}, the kurtosis 
does not depend on $\phi_0$ for long times. The long-time tail, being proportional to $1/t$, is more pronounced than that for the skewness.
Moreover, as displayed in figures \ref{fig:Woelbung} and \ref{fig:Woelbunga2}, for initial angles $\phi_0 \neq 0.5\pi$ 
the distribution is leptokurtic (positive kurtosis) for relatively short times and platykurtic (negative kurtosis) for 
relatively long times. Thus for intermediate times a change of sign is induced such that the kurtosis approaches its
asymptotic value zero from below. This is in contrast to 
passive ellipsoidal particles in two dimensions~\cite{Han:06} where non-Gaussian behaviour is
due to dissipatively coupled translational and rotational motion. In the latter case, the same scaling of the long-time tail
proportional to $1/t$ is found for the kurtosis but it approaches zero from above. 

We expect that the different sign is linked to the one-dimensionality of
our model rather than to the qualitatively different translation-rotation
coupling, which is due to the driving force in our model as opposed to the
different transverse and parallel short-time translational diffusivities
in the passive ellipsoidal particle model. In particular, we expect the negative kurtosis at long times $t\gg D_r^{-1}$ to reflect a broad translational van Hove function~\cite{hansen-mcdonald:86} with shorter tails as compared to a Gaussian distribution, which is attributed to the non-linear $\cos$-term in equation (\ref{Langevinx1}). 

\section{Results for a time-dependent torque}
\label{+torque}
\begin{figure}[htb]
\centerline{\includegraphics[width=0.65\textwidth]{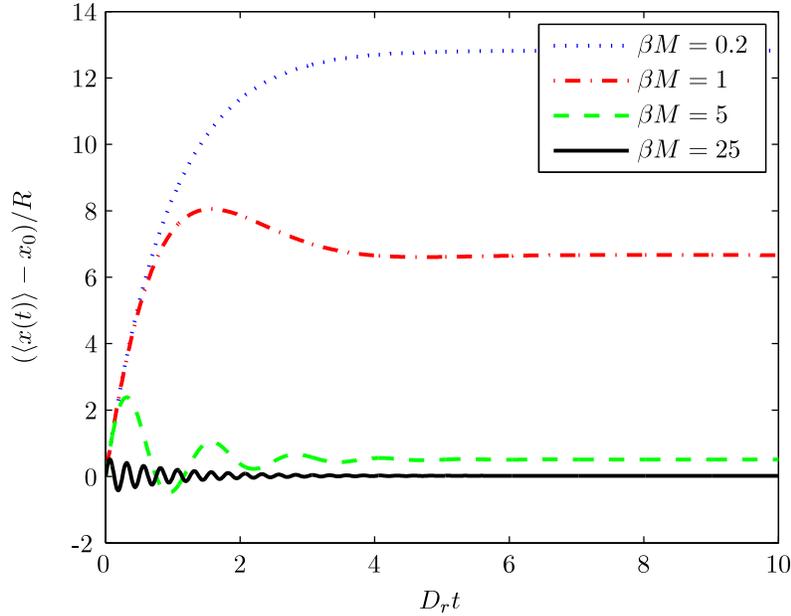}}
\caption{Mean position of a spherical particle with additional constant external torque for $\beta RF_0=10$, $\phi_0=0$ and different values of $\beta M$.}
\label{fig:Moment12}
\end{figure}

Let us now assume an additional internal or external torque. Before considering the case of an arbitrarily time-dependent torque $M(t)$, we first consider a constant torque $M$. Solving the Langevin equations (\ref{Langevinx1}) and (\ref{Langevinphi1}) under this assumption, one obtains 
\begin{equation}
\label{Mittelwertphi2}
\langle \phi(t)\rangle = \phi_0 +\beta D_\mathrm{r} M t = \phi_0 + \omega t
\end{equation}
with the frequency $\omega = \beta D_\mathrm{r}M$ and 
\begin{equation}
\left\langle (\phi(t)-\langle \phi(t)\rangle )^2\right\rangle =2D_\mathrm{r}t.
\end{equation}
By replacing $\phi_0$ in formula (\ref{Verteilungphikonkret}) with $\phi_0 + \omega t$, the updated probability distribution of $\phi$ is gained. The mean position is obtained as 
\begin{eqnarray}
\label{Moment12}
\langle x(t)-x_0 \rangle & = & \frac{\beta D}{(D_\mathrm{r}^2+\omega^2)}F_0\bigl[D_\mathrm{r}\cos(\phi_0)-\omega \sin(\phi_0) \nonumber \\
& & +e^{-D_\mathrm{r}t}(\omega\sin(\phi_0+\omega t)-D_\mathrm{r}\cos(\phi_0+\omega t))\bigr].
\end{eqnarray}
In figure \ref{fig:Moment12} this result is plotted for different values of the dimensionless quantity $\beta M$, which is the ratio of the external torque over the thermal energy. The long-time mean position is given by
\begin{equation}
\lim_{t \to \infty} \langle x(t)-x_0\rangle =  \frac{\beta D}{(D_\mathrm{r}^2+\omega^2)}F_0\left[D_\mathrm{r}\cos(\phi_0)-\omega \sin(\phi_0) \right],
\end{equation}
while the behaviour for short times is the same as in formula (\ref{Mittelwerts}) for a vanishing torque. 

\begin{figure}[htb]
\centerline{\includegraphics[width=0.65\textwidth]{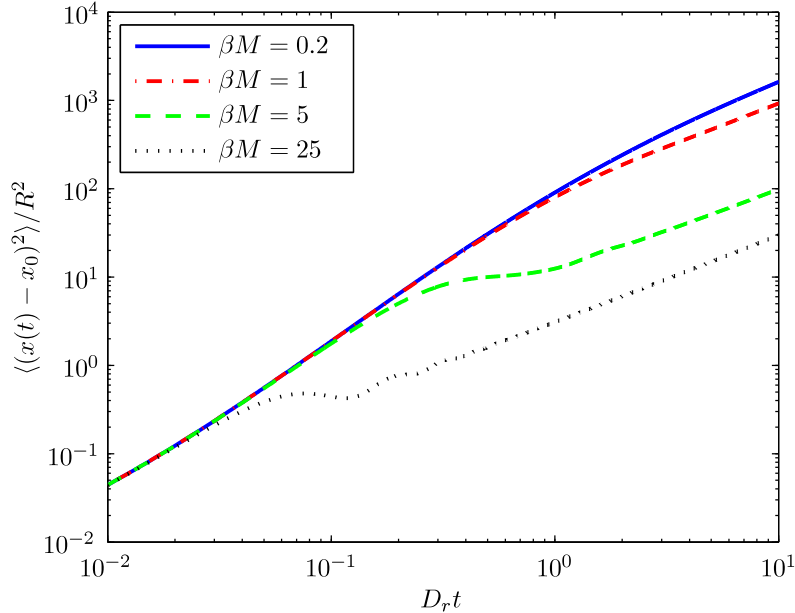}}
\caption{Mean square displacement of a spherical particle with additional constant external torque for $\beta RF_0=10$, $\phi_0=0$ and different values of $\beta M$.}
\label{fig:Moment22}
\end{figure}

Following the notation introduced in formula (\ref{Korrelation1}) the Green function is now given by 
\begin{equation}
\label{Green2}
G(\phi_1,\phi_2 ,t_1-t_2)=\frac{1}{\sqrt{4\pi D_\mathrm{r}(t_1-t_2)}}\exp\left({-\frac{(\phi_1-\phi_2 -\omega (t_1-t_2))^2}{4D_\mathrm{r}(t_1-t_2)}}\right).
\end{equation}
This leads to 
\begin{equation}
\label{Korrelation2a}
\left\langle \cos\phi_1 \cos\phi_2\right\rangle_{t_1>t_2} = \frac{1}{2}e^{-D_\mathrm{r}(t_1-t_2)}\left[\cos(\omega (t_1-t_2))+\cos\left(2\phi_0 +\omega (t_1+t_2)\right)e^{-4D_\mathrm{r}t_2}\right]
\end{equation}
and by integration one obtains
\begin{eqnarray}
\label{Moment22}
\left\langle (x(t)-x_0)^2\right\rangle\!\!\! &\!=\!\!\!& 2Dt + \beta^2 F_0^2 D^2\biggl\{\frac{D_\mathrm{r}t}{D_\mathrm{r}^2+\omega^2}-\frac{D_\mathrm{r}^2-\omega^2}{(D_\mathrm{r}^2+\omega^2)^2}
\nonumber \\
 & & -\frac{e^{-D_\mathrm{r}t}}{(D_\mathrm{r}^2+\omega^2)^2}\Bigl[(\omega^2-D_\mathrm{r}^2)\cos(\omega t)+2\omega D_\mathrm{r}\sin(\omega t)\Bigr]
\nonumber \\
& & +\frac{1}{(9D_\mathrm{r}^2 +\omega^2)(D_\mathrm{r}^2+\omega^2)}\Bigl[e^{-D_\mathrm{r}t}\bigl((-3D_\mathrm{r}^2+\omega^2)\cos(2\phi_0+\omega t)
\nonumber \\
& & +4D_\mathrm{r}\omega\sin(2\phi_0+\omega)\bigr)- (-3D_\mathrm{r}^2+\omega^2)\cos(2\phi_0)-4D_\mathrm{r}\omega\sin(2\phi_0)\Bigr] 
\nonumber \\
& & +\frac{1}{(9D_\mathrm{r}^2+\omega^2)(16D_\mathrm{r}^2+4\omega^2)}\Bigl[e^{-4D_\mathrm{r}t}\bigl((12D_\mathrm{r}^2-2\omega^2) \cos(2\phi_0+2\omega t)
\nonumber \\
& & - 10 D_\mathrm{r} \omega \sin(2\phi_0+2\omega t)\bigr)- (12D_\mathrm{r}^2-2 \omega^2)\cos(2\phi_0)+10 D_\mathrm{r}\omega\sin(2\phi_0)\Bigr]\!\!\biggr\}.
\end{eqnarray}
The result is displayed in figure \ref{fig:Moment22}. In this case, the long-time diffusion coefficient is given by 
\begin{equation}
D_\mathrm{l}=D+\frac{8}{9} \frac{(\beta F_0R^2)^2D_\mathrm{r}}{(1+(\beta M)^2)}. 
\end{equation} 

To generalize the preceding considerations, the torque $M(t)$ is assumed to be arbitrarily time-dependent now. Similarly to the two special cases investigated so far, it can be seen that the mean position of the particle is given by 
\begin{equation}
\label{Moment1x3}
\langle x(t)\rangle =\beta F_0D\int_0^t\cos\left[\phi_0+\beta D_\mathrm{r}\int_0^{t_1}M(t_2)\mathrm{d}t_2\right]e^{-D_\mathrm{r}t_1}\mathrm{d}t_1+x_0.
\end{equation} 
The calculation of the mean square displacement starts with formula (\ref{Moment2x1}) again. The first summand is the most interesting one because the other ones can be treated as before. Based on the formula
\begin{eqnarray}
\label{Korrelationt}
\left\langle \cos\phi_1 \cos\phi_2\right\rangle_{t_1>t_2} &=& \frac{1}{2}e^{-D_\mathrm{r}(t_1-t_2)}\Bigl[\cos\left(\beta D_\mathrm{r}\int_{t_2}^{t_1}M(t)\mathrm{d}t\right) 
\nonumber \\
&&+ \cos\left(2\phi_0+ 2\beta D_\mathrm{r}\int_{0}^{t_2}M(t)\mathrm{d}t + \beta D_\mathrm{r}\int_{t_2}^{t_1}M(t)\mathrm{d}t\right) e^{-4D_\mathrm{r}t_2}\Bigr]
\end{eqnarray}
we introduce
\begin{eqnarray}
\omega_{t_1}\!&\!:=\!&\!\beta D_\mathrm{r} \int_0^{t_1}M(t)\mathrm{d}t, \nonumber \\
\omega_{t_2}\!&\!:=\!&\!\beta D_\mathrm{r} \int_0^{t_2}M(t)\mathrm{d}t.
\end{eqnarray}
Using this notation the problem can  be solved in a similar way as for a constant $M$. The mean square displacement is now given by
\begin{eqnarray}
\label{Moment2xallgemein}
\left\langle(x(t)-x_0)^2\right\rangle &=& 2Dt + \beta^2 F_0^2D^2\int_0^t\mathrm{d}t_1\int_0^{t_1}\mathrm{d}t_2 e^{-D_\mathrm{r}(t_1-t_2)}
\nonumber \\
&& \times \left[\cos(\omega_{t_1}-\omega_{t_2}) + \cos\left(2\phi_0 + \omega_{t_1} + \omega_{t_2}\right) e^{-4D_\mathrm{r}t_2}\right].
\end{eqnarray}

\section{Conclusions}
\label{conclusions}
In conclusion, motivated by recent experiments on catalytic colloidal 
particles~\cite{Baraban:08,Baraban2,Baraban_private},
we have proposed and solved a model for a self-propelled colloidal particle on a substrate.
An internal or external time-dependent torque is also included in the most general version of the model
which can arise, e.g., from an external magnetic field.
The first four moments of the particle displacement distribution were calculated analytically.
Significant non-Gaussian behaviour was found for intermediate time.
The normalized kurtosis changes sign and approaches zero from below with a massive long-time tail
inversely proportional to time.

Future work should address several generalizations of the model. First of all,
the one-dimensionality of our model can be generalized towards higher 
dimensions both for the translational and orientational degrees of freedom. 
In particular, the translational degrees of freedom can be considered to be two-dimensional (in a plane), 
and the orientational ones on a sphere. For the latter case, first analytical results have been obtained~\cite{tenHagen}. Also, e.g., for weak gravity, the third translational dimension perpendicular
to the substrate is getting important, which  results in unusual sedimentation effects~\cite{Poon}.
Furthermore, the self-propelled particle can be confined in the lateral direction~\cite{Baraban_private}
which leads to a finite mean square displacement. This effect should  be incorporated into a model study as well.
First results have been obtained for a circle-swimmer in planar circular geometry~\cite{Zimmermann}
and for swimmers in cuspy environments leading to self-rotating objects~\cite{Leonardo}.

Last not least, the collective behaviour of many interacting self-propelled particles is expected to lead to novel
characteristic nonequilibrium effects both without~\cite{Viczek:PRL,EPL_Raina,Markus_Baer,Romanczuk:09} and with confinement~\cite{Leonardo,Wensink}.
As stated in the introduction, the Smoluchowski equation,  suitably generalized to self-propelled particles~\cite{Wensink},
is an appropriate starting point here and the 
general hierarchy of Bogolyubov-Born-Green-Kirkwood-Yvon~\cite{Bogolyubov1,Bogolyubov2,Uhlenbeck} is expected to be a valuable tool in order to derive
approximations in a systematic way. This fact after all clearly
links the present paper to the 100th anniversary of N. N. Bogolyubov.

\section*{Acknowledgements}

We thank L. Baraban, A. Erbe and P. Leiderer for helpful discussions which have stimulated the study of our model.
We further thank H. H. Wensink and U. Zimmermann  for helpful suggestions.  This work has been supported by
the DFG through the SFB TR6. We dedicate this work to the 100th anniversary of N. N. Bogolyubov.

\appendix
\section*{Appendix}
Using the notation $a=\beta RF_0$ and a scaled time 
$\tau =D_\mathrm{r}t$, we summarize here the analytical results for the third and fourth moments
as well as for the  skewness $S$ and kurtosis $\gamma$:
\begin{eqnarray}
\left\langle \frac{(x(t)-x_0)^3}{R^3}\right\rangle & = & {\frac {32}{3}}\,a\tau\cos \left( \phi_0 \right)  \left( 1-{e^{-\tau}} \right) \nonumber \\ 
&& +{\frac {64}{27}}\,{a}^{3} \Bigl( -{\frac {45}{8}}\,\cos \left( \phi_0
 \right) +\frac{1}{24}\,\cos \left( 3\,\phi_0 \right) +{\frac {17}{3}}\,\cos
 \left( \phi_0 \right) {e^{-\tau}}\nonumber \\ 
 &&- \frac{1}{16}\,\cos \left( 3\,\phi_0 \right) {e^{-\tau}}
 +3\,\cos \left( \phi_0 \right) \tau-\frac{1}{24}\,\cos \left( \phi_0 \right) {e^{-4\,\tau}}\nonumber \\ 
 &&+\frac{1}{40}\,\cos \left( 3\,\phi_0 \right) {e^{-4\,\tau}}+\frac{5}{2}\,\cos \left( \phi_0 \right) \tau{e^{-\tau}}-  {\frac {1}{240}}\,{e^{-9\,\tau}}\cos \left( 3\,\phi_0 \right)  \Bigr), \qquad
\end{eqnarray}
\begin{eqnarray}
\left\langle \frac{(x(t)-x_0)^4}{R^4}\right\rangle & = & {\frac {64}{3}}\,{\tau}^{2}+{\frac {256}{9}}\,{a}^{2}\tau \left( {e^{-\tau}}+\tau-
1+\frac{1}{12}\,\cos \left( 2\,b \right)  \left( {e^{-4\,\tau}}-4\,{e^{-\tau}}+3
 \right)  \right) \nonumber \\ 
 &&+{\frac {256}{81}}\,{a}^{4}\biggl(3\,{\tau}^{2}+{\frac {1}{6720}}\,{e^{-16\,\tau}}\cos \left( 4\,\phi_0 \right) -5\,\tau{e^{-\tau}}-
{\frac {45}{4}}\,\tau \nonumber \\ 
 &&+{\frac {261}{16}}+{\frac {1}{600}}\,\cos \left( 2\,
\phi_0 \right) {e^{-9\,\tau}}-{\frac {19}{6}}\,\cos \left( 2\,\phi_0 \right) +{
\frac {1}{240}}\,\cos \left( 4\,\phi_0 \right) {e^{-4\,\tau}}\nonumber \\ 
 &&+{\frac {1}{192}}
\,\cos \left( 4\,\phi_0 \right) +\frac{1}{48}\,{e^{-4\,\tau}}-{\frac {49}{3}}\,{e^{-\tau}
}-{\frac {7}{450}}\,\cos \left( 2\,\phi_0 \right) {e^{-4\,\tau}}\nonumber \\ 
 &&-{\frac {1}{
120}}\,\cos \left( 4\,\phi_0 \right) {e^{-\tau}}+\frac{3}{2}\,\tau\cos \left( 2\,\phi_0
 \right) -\frac{1}{30}\,\tau\cos \left( 2\,\phi_0 \right) {e^{-4\,\tau}}\nonumber \\ 
 &&+{\frac {229}{72}
}\,\cos \left( 2\,\phi_0 \right) {e^{-\tau}}-{\frac {1}{840}}\,\cos \left( 4\,
\phi_0 \right) {e^{-9\,\tau}}+\frac{5}{3}\,\tau\cos \left( 2\,\phi_0 \right) {e^{-\tau}}\biggr),\qquad
\end{eqnarray}
\begin{eqnarray}
\label{Schiefe}
S&=& \biggl[ \frac{8}{3}\,\tau+{\frac {16}{9}}\,{a}^{2} \left( {e^{-\tau}}+\tau-1+\frac{1}{12}\,\cos
 \left( 2\,\phi_0 \right)  \left( {e^{-4\,\tau}}-4\,{e^{-\tau}}+3 \right) 
 \right) \nonumber \\
&& -{\frac {16}{9}}\,{a}^{2} \left( \cos \left( \phi_0 \right) 
 \right) ^{2} \left( 1-{e^{-\tau}} \right) ^{2} \biggr] ^{-3/2} \nonumber \\
&&\times \Biggl\{-{\frac {32}{9}}\,{a}^{3}\cos \left( \phi_0 \right) -{\frac {760}{81}}\,{a}
^{3}\cos \left( \phi_0 \right) {e^{-\tau}}-{\frac {44}{27}}\,{a}^{3}\cos
 \left( 3\,\phi_0 \right) {e^{-\tau}}-{\frac {32}{81}}\,{a}^{3}\cos \left( \phi_0
 \right) {e^{-4\,\tau}}\nonumber \\
&&-{\frac {32}{135}}\,{a}^{3}\cos \left( 3\,\phi_0
 \right) {e^{-4\,\tau}}-{\frac {4}{405}}\,{a}^{3}{e^{-9\,\tau}}\cos \left( 3
\,\phi_0 \right) +{\frac {32}{81}}\,{a}^{3}\cos \left( 3\,\phi_0 \right) +{
\frac {352}{27}}\,{a}^{3}\cos \left( \phi_0 \right) \tau{e^{-\tau}}\nonumber \\
&&+{\frac {448}{
27}}\,{a}^{3}\cos \left( \phi_0 \right) {e^{-2\,\tau}}+{\frac {64}{27}}\,{a}^{
3}{e^{-2\,\tau}}\cos \left( 3\,\phi_0 \right) -{\frac {32}{27}}\,{a}^{3}{e^{-3
\,\tau}}\cos \left( 3\,\phi_0 \right) -{\frac {32}{9}}\,{a}^{3}{e^{-3\,\tau}}\cos
 \left( \phi_0 \right)\nonumber \\
&& +{\frac {8}{27}}\,{a}^{3}{e^{-5\,\tau}}\cos \left( \phi_0
 \right) +{\frac {8}{27}}\,{a}^{3}{e^{-5\,\tau}}\cos \left( 3\,\phi_0 \right)
\Biggr\}
\end{eqnarray}
and 
\begin{eqnarray}
\label{kurtosis}
\gamma&=& \biggl[ \frac{8}{3}\,\tau+{\frac {16}{9}}\,{a}^{2} \left( {e^{-\tau}}+\tau-1+\frac{1}{12}\,\cos
 \left( 2\,\phi_0 \right)  \left( {e^{-4\,\tau}}-4\,{e^{-\tau}}+3 \right) 
 \right) \nonumber \\
 && -{\frac {16}{9}}\,{a}^{2} \left( \cos \left( \phi_0 \right) 
 \right) ^{2} \left( 1-{e^{-\tau}} \right) ^{2} \biggr] ^{-2} \nonumber \\
&& \times \Biggl\{-{\frac {128}{9}}\,\tau{a}^{2}{e^{-2\,\tau}}\cos ( 2\,\phi_0 ) +{
\frac {64}{81}}\,{a}^{4}{e^{-6\,\tau}}\cos ( 2\,\phi_0 ) +{\frac {
64}{3}}\,{\tau}^{2}-{\frac {15424}{729}}\,{a}^{4}{e^{-\tau}}\cos ( 2\,\phi_0
 ) \nonumber \\ 
 &&+{\frac {32}{81}}\,{a}^{4}{e^{-6\,\tau}}\cos ( 4\,\phi_0
 ) +{\frac {64}{45}}\,{a}^{4}{e^{-\tau}}\cos ( 4\,\phi_0 ) +{
\frac {2032}{27}}\,{a}^{4}-{\frac {128}{3}}\,{a}^{2}\tau \nonumber \\
&&+{\frac {256}{9}}
\,{a}^{2}{\tau}^{2}-{\frac {1216}{27}}\,{a}^{4}\tau+{\frac {256}{27}}\,{a}^{
4}{\tau}^{2}+{\frac {64}{27}}\,\tau{a}^{2}{e^{-4\,\tau}}\cos ( 2\,\phi_0
 ) -{\frac {70976}{18225}}\,{a}^{4}{e^{-4\,\tau}}\cos ( 2\,\phi_0
 ) \nonumber \\
  &&-{\frac {20480}{243}}\,{a}^{4}{e^{-\tau}}-{\frac {688}{243}}\,{a
}^{4}{e^{-4\,\tau}}-{\frac {2560}{81}}\,{a}^{4}\tau{e^{-\tau}}+{\frac {512}{9}}
\,\tau{a}^{2}{e^{-\tau}}-{\frac {128}{9}}\,\tau{a}^{2}{e^{-2\,\tau}}\nonumber \\
&&+{\frac {2048}
{81}}\,{a}^{4}\tau{e^{-2\,\tau}}  -{\frac {1136}{1215}}\,{a}^{4}{e^{-4\,\tau}}
\cos ( 4\,\phi_0 ) -{\frac {256}{81}}\,{a}^{4}{e^{-2\,\tau}}\cos
 ( 4\,\phi_0 ) +{\frac {32}{81}}\,{a}^{4}{e^{-6\,\tau}}\nonumber \\
&&-{\frac {
2560}{243}}\,{a}^{4}\tau{e^{-\tau}}\cos ( 2\,\phi_0 )  +{\frac {4}{8505
}}\,{a}^{4}{e^{-16\,\tau}}\cos ( 4\,\phi_0 ) 
 -{\frac {256}{243}}\,{
a}^{4}{e^{-5\,\tau}}-{\frac {2048}{1215}}\,{a}^{4}{e^{-5\,\tau}}\cos ( 
2\,\phi_0 )\nonumber \\
 && +{\frac {2048}{81}}\,{a}^{4}\tau{e^{-2\,\tau}}\cos ( 2\,\phi_0
 ) +{\frac {1792}{81}}\,{a}^{4}{e^{-3\,\tau}}+{\frac {1088}{81}}\,{
a}^{4}\cos ( 2\,\phi_0 ) -{\frac {20}{81}}\,{a}^{4}\cos ( 4
\,\phi_0 )\nonumber \\
 && +{\frac {2048}{81}}\,{a}^{4}{e^{-3\,\tau}}\cos ( 2\,\phi_0
 ) +{\frac {256}{81}}\,{a}^{4}{e^{-3\,\tau}}\cos ( 4\,\phi_0
 ) -{\frac {128}{27}}\,{a}^{4}\tau\cos ( 2\,\phi_0 ) -{\frac 
{64}{9}}\,{a}^{2}\tau\cos ( 2\,\phi_0 )\nonumber \\
&& -{\frac {2336}{243}}\,{a}^{
4}{e^{-2\,\tau}} -{\frac {32}{1215}}\,{a}^{4}{e^{-10\,\tau}}\cos ( 2\,\phi_0
 ) -{\frac {32}{1215}}\,{a}^{4}{e^{-10\,\tau}}\cos ( 4\,\phi_0
 ) -{\frac {3104}{243}}\,{a}^{4}{e^{-2\,\tau}}\cos ( 2\,\phi_0
 ) \nonumber \\
 && +{\frac {64}{2025}}\,{a}^{4}{e^{-9\,\tau}}\cos ( 2\,\phi_0
 ) -{\frac {256}{405}}\,{a}^{4}{e^{-5\,\tau}}\cos ( 4\,\phi_0
 ) +{\frac {512}{27}}\,{a}^{2}\tau{e^{-\tau}}\cos ( 2\,\phi_0 ) \nonumber \\
&& -{\frac {128}{1215}}\,{a}^{4}\tau{e^{-4\,\tau}}\cos ( 2\,\phi_0 ) +{
\frac {64}{2835}}\,{a}^{4}{e^{-9\,\tau}}\cos ( 4\,\phi_0 )\Biggr\}-3.
\end{eqnarray}

\label{last@page}
\end{document}